\documentclass[aps,prb,a4paper,twocolumn,floatfix,showpacs,preprintnumbers]{revtex4-1}
\usepackage{amsmath}
\usepackage{amssymb}
\usepackage{graphicx}
\usepackage{color}

\begin{document}

\title{Spin polarization and $g$-factor enhancement in graphene nanoribbons
in magnetic field}
\author{S. Ihnatsenka}
\affiliation{Department of Physics, Simon Fraser University, Burnaby, British Columbia,
Canada V5A 1S6}
\email{sihnatse@sfu.ca}
\author{I. V. Zozoulenko}
\affiliation{Solid State Electronics, ITN, Link\"{o}ping University, 601 74, Norrk\"{o}%
ping, Sweden}
\email{igozo@itn.liu.se}

\begin{abstract}
We provide a systematic quantitative description of spin polarization in
armchair and zigzag graphene nanoribbons in a perpendicular magnetic field.
We first address spinless electrons within the Hartree approximation
studying the evolution of the magnetoband structure and formation of the
compressible strips. We discuss the potential profile and the density
distribution near the edges and the difference and similarities between
armchair and zigzag edges. Accounting for the Zeeman interaction and
describing the spin effects via the Hubbard term we study the spin-resolved
subband structure and relate the spin polarization of the system at hand to
the formation of the compressible strips for the case of spinless electrons.
At high magnetic field the calculated effective $g$-factor varies around a
value of $\left\langle g^{\ast }\right\rangle \approx 2.25$ for armchair
nanoribbons and $\left\langle g^{\ast }\right\rangle \approx 3$ for zigzag
nanoribbons. An important finding is that in zigzag nanoribbons the
zero-energy mode remains pinned to the Fermi-energy and becomes fully
spin-polarized for all magnetic fields, which, in turn, leads to a strong
spin polarization of the electron density near the zigzag edge.
\end{abstract}

\date{\today }
\pacs{72.80.Vp, 73.22.Pr, 73.63.Nm, 73.43.-f}
\maketitle

\section{Introduction}

Investigation of effects of electron interaction and spin in graphene at
high magnetic field represents one of the frontiers in the graphene
research. Even though many aspects of the magnetoconductance of graphene
related to the formation of unconventional Landau level spectra and the
anomalous Hall effect are well understood theoretically and confirmed
experimentally\cite{Castro_Neto_review,GusyninQHE,Novoselov2005,Zhang},
there are still a number of questions awaiting their resolution. One of
these questions which is extensively debated in the current literature is
the origin of the splitting of the lowest Landau level and the emerging of
insulating state at the Dirac point.\cite%
{Zhang2006,Jiang,Checkelsky,Zhang2009,Zhao12} Even though the precise origin
of this state is under current debate, it is generally believed that it is
related to electron-electron interaction and spin effects. The importance of
electron interaction was also outlined for higher Landau levels \cite{Jiang}%
. Recently, spin-splitting in graphene and bilayer graphene in high magnetic
field was experimentally analyzed by Kurganona \textit{et al.}\cite%
{Kurganova}, who found that $g$-factor in graphene is enhanced, and
attributed this to electron-electron interaction effects. The spin-splitting
of the states in graphene\cite{Folk} and graphene quantum dots\cite%
{Guttinger} was also studied in a parallel magnetic filed.

Motivated by this interest to the electron interacton and spin effects in
graphene in the high magnetic field, in the present paper we study the spin
polarization and enhancement of the $g$-factor in graphene nanoribbons
(GNRs). Note that various aspects of electron and spin interactions in high
magnetic have been extensively studied in conventional semiconducting
quantum wires defined in two dimensional electron gas (2DEG) \cite%
{Kinaret,Dempsey,Tokura,takis2002,Stoof,Ihnatsenka_wire1,Ihnatsenka_wire_comp_strips,IhnatsenkaCEOQW,IhnatsenkaMarcus}%
. One of the motivations for such studies is related to advances in
semiconductor spintronics utilizing the spin degree of freedom for adding
new functionalities to electronic devices.\cite{spintronics} Some of
proposed and investigated devices for spintronics and quantum computation
applications operates in the edge state regime,\cite{adot,Giovannetti} which
obviously requires a detailed knowledge of the structure of the states in a
quantum wire or at the edge of the 2DEG. The properties and detailed
information about propagating states at the boundaries are also essential
for interpretation of experiments in various electron interferometers in the
quantum Hall regime\cite{Camino,IhnatsenkaKirzcenow,Paradiso}. Because
graphene represents a very promising system for implementation of many
devices and concepts for spintronics and quantum information processing
applications utilizing the edge state transport regime a detailed knowledge
of the density and potential profiles near the edges as well as spin
properties are important for understanding and designing of such devices.

The paper is organized as follows. In Sec. II we present a formulation of
the problem, define the Hamiltonian and briefly outline the self-consistent
computational scheme. The results and discussion are presented in Sec. II.
Section IIA discusses the potential profile and the charge accumulation near
the edges in ribbons of different widths and edge terminations. Section IIB
is devoted to the case of spinless electrons focussing on the formation of
compressible strips and evolution of the magnetoband structure. Finally,
based on the results of Sec. IIB, Sec. IIC discusses the spin splitting and
the enhancement of the $g$-factor for the case of electrons with spin. The
conclusions of the work are presented in Sec. III.

\section{Model}

We consider an infinite GNR of the width $W$, located in an insulating
substrate with the relative permittivity $\epsilon _{r}$ and subjected to
the perpendicular magnetic field $B$, see inset to Fig. \ref{fig:edge}(b). A
metallic back gate situated at the distance $d$ from the ribbon is used to
tune the Fermi energy in order to change an electron concentration in the
GNR. The system is modeled by the \textit{p}-orbital tight-binding
Hubbard-type Hamiltonian in the mean-field approximation, $H=H^{\uparrow
}+H^{\downarrow },$ which is shown to describe carbon electron systems in
good agreement with the first-principles calculations\cite{Palacios, Yazyev},%
\begin{align}
& H^{\sigma }=-\sum_{\mathbf{r},\Delta }t_{\mathbf{r},\mathbf{r}+\Delta }a_{%
\mathbf{r}\sigma }^{+}a_{\mathbf{r}+\Delta ,\sigma }  \label{H} \\
& +\sum_{\mathbf{r}}\left( V_{Z}^{\sigma }+V_{H}(\mathbf{r})+V_{U}^{\sigma
^{\prime }}(\mathbf{r})\right) a_{\mathbf{r}\sigma }^{+}a_{\mathbf{r}\sigma }
\notag
\end{align}%
where $\sigma $, $\sigma ^{\prime }$ correspond to two opposite spin states $%
\uparrow $, $\downarrow $; the summation runs over all sites $\mathbf{r=(}x,y%
\mathbf{)}$ of the graphene lattice, $\Delta $ includes the nearest
neighbors only. The magnetic field is included in a standard way via the
Pierel's substitution, $t_{\mathbf{r},\mathbf{r}+\Delta }=t_{0}\exp (i2\pi
\phi _{\mathbf{r},\mathbf{r}+\Delta }/\phi _{0})$ where $\phi _{\mathbf{r},%
\mathbf{r}+\Delta }=\int_{\mathbf{r}}^{\mathbf{r}+\Delta }\mathbf{A}\cdot d%
\mathbf{l},$ with $\mathbf{A}$ being the vector potential, and $\phi _{0}=h/e
$ being the magnetic flux quantum, $t_{0}=2.7$ eV. (In our calculations we
use the Landau gauge, $\mathbf{A}=(-By,0)$). The first two terms in Eq. (\ref%
{H}) correspond to the non-interacting part of the Hamiltonian, with the
first term describing the kinetic energy of the electrons on a graphene
lattice. The second term describes the Zeeman energy triggering the
spin-splitting in the magnetic field, $V_{Z}^{\sigma }=\pm \frac{1}{2}g\mu
_{b}B,$ where $+(-)$ signs correspond to the opposite spin states $\uparrow $
$(\downarrow );$ $g=2$ is the bare $g$-factor of pristine graphene, and the
Bohr magneton $\mu _{b}=e\hbar /2m_{e}.$ The two last terms in Eq. (\ref{H})
describe the electron interaction. The long-range Coulomb interaction
between induced charges in the GNR is given by the standard Hartree term,
\begin{equation}
V_{H}(\mathbf{r})=\frac{e^{2}}{4\pi \varepsilon _{0}\varepsilon _{r}}\sum_{%
\mathbf{r}^{\prime }\neq \mathbf{r}}n_{\mathbf{r}^{\prime }}\left( \frac{1}{|%
\mathbf{r}-\mathbf{r}^{\prime }|}-\frac{1}{\sqrt{|\mathbf{r}-\mathbf{r}%
^{\prime }|^{2}+4d^{2}}}\right) ,  \label{VH}
\end{equation}%
where $n_{\mathbf{r}}=n_{\mathbf{r}\uparrow }+n_{\mathbf{r}\downarrow }$ is
the total electron density, and the second term corresponds to a
contribution from the mirror charges. The last term in the Hamiltonian (\ref%
{H}) corresponds to the Hubbard energy,%
\begin{equation}
V_{U}^{\sigma ^{\prime }}(\mathbf{r})=U\left( n_{\mathbf{r}}^{\sigma
^{\prime }}-\frac{1}{2}\right) ,  \label{U}
\end{equation}%
and describes repulsion between electrons of the opposite spins on the same
site. The number of excess electron at the site $\mathbf{r}$ reads,
\begin{equation}
n_{\mathbf{r}}^{\sigma }=\int_{-\infty }^{\infty }\rho ^{\sigma }(\mathbf{r}%
,E)f_{FD}(E,E_{F})dE-n_{\text{ions}},  \label{density}
\end{equation}%
where $\rho ^{\sigma }(\mathbf{r},E)=-\frac{1}{\pi }\Im \lbrack G^{\sigma }(%
\mathbf{r},\mathbf{r},E)]$ is the energy dependent local density of states
(LDOS) at rero temperature; $G^{\sigma }(\mathbf{r},\mathbf{r},E)$ is the
Green's function in the real space representation of an electron of the spin
$\sigma $ residing on the site $\mathbf{r}$, $E_{F}=eV_{g}$ is the Fermi
energy which value is adjusted by the gate voltage, and $n_{\text{ions}%
}=1/A_{0}=3.8\times 10^{19}$m$^{-2}$ is the positive charge background of
ions ($A_{0}=\frac{3\sqrt{3}}{4}a_{0}^{2}$ is the area per one C atom and $%
a_{0}=0.142$ nm is the C-C distance). Equations (\ref{H})-(\ref{density})
are solved self-consistently using the Green's function technique in order
to calculate the band structure, the charge density and the potential
distribution.\cite{Xu,Shylau10,Shylau11} For a given potential distribution
we compute the conductance using the Landauer formula
\begin{equation}
\mathcal{G}^{\sigma }(E_{F})=\frac{e^{2}}{h}\int T^{\sigma }(E)\left[ -\frac{%
\partial f_{FD}(E-E_{F})}{\partial E}\right] dE,  \label{G}
\end{equation}%
where $T^{\sigma }(E)$ is the total transmission coefficient for electrons
with spin $\sigma $, and $f_{FD}$ is the Fermi-Dirac distribution.

\section{Results and discussion}

\subsection{Potential profile and the charge accumulation near the edges in
ribbons of different widths and edge terminations.}

%*********************************************************
\begin{figure}[tbh]
\includegraphics[keepaspectratio,width=\columnwidth]{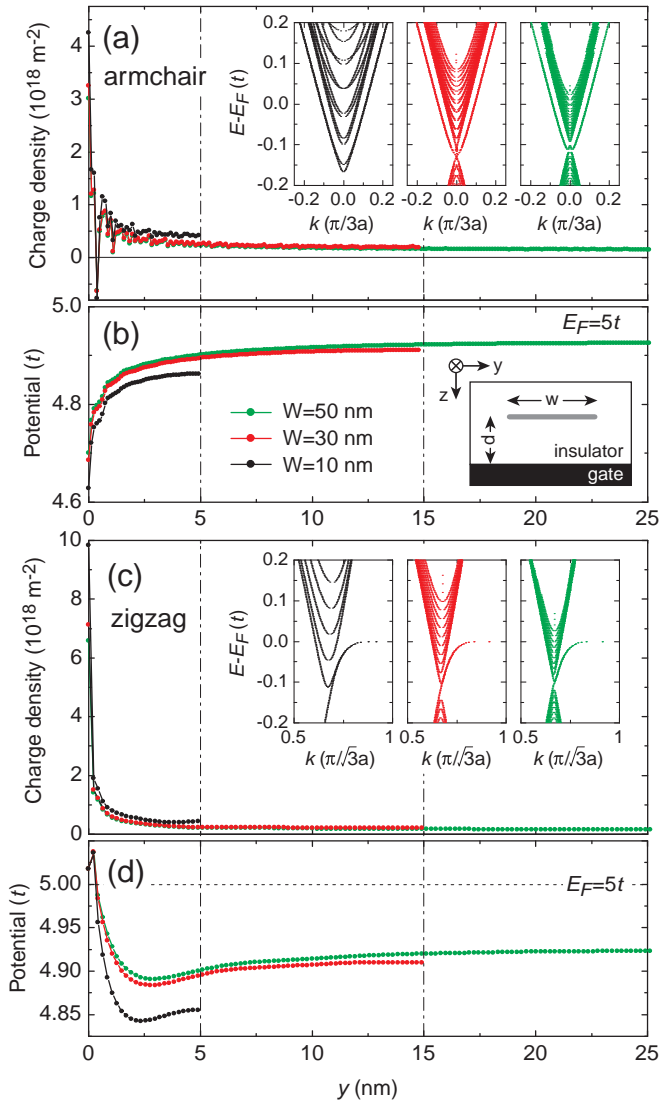}
\caption{(Color online). (a),(c) The self-consistent charge densities and
(b),(d) potentials for the armchair and zigzag ribbons of widths $W$=10,30
and 50 nm calculated in the Hartree approximation at zero magnetic field.
Only half of the ribbon is shown. The insets in (a),(c) show the dispersion
relation for nanoribbons of different widths $W$=10, 30 and 50 nm from the
left to the right for several lowest subbands.The inset in (b) shows a
schematically geometry of the device; $d=30$ nm. $E_F$=5 $t$. $t=2.7$ eV. $%
a=0.142$ nm.}
\label{fig:edge}
\end{figure}
%*********************************************************

In the present study we aim at the description of spin polarization in
realistically wide GNRs. In current experiments\cite%
{Molitor,Oostinga,Poumirol,Ribeiro} the widths of nanoribbons are $W\sim 100$
nm and the magnetic field reaches $B\lesssim 60$ T, corresponding to the
ratio $W/l_{B}\approx 10$, with $l_{B}=\sqrt{\hbar /eB}$ being the magnetic
length. At the same time, because of computational limitations, it is
difficult to treat ribbons of widths exceeding 50 nm. Therefore in our
calculations we re-scale the system using ribbons of smaller width $W\sim 30$
nm subjected to higher fields (up to $B\sim 350$ T) in order to keep the
ratio $W/l_{B}\approx 10$ in accordance with typical experiments.\cite%
{Molitor,Oostinga,Poumirol,Ribeiro} In very narrow ribbons the quantum
confinement effects can dominate ribbon's electronic properties. Thus, a
concern might arise whether the obtained results remain valid for
realistically wide ribbons. In the present section we investigate how
nanoribbon's electronic properties such as the density distribution and the
potential profile in the vicinity of the edge change with the increase of
the ribbon's width, and find that the width of $W\approx 30$ nm is already
sufficient to capture all essential features of a wide ribbon or even a
semi-infinite graphene sheet.

In our study we consider both types of edges, armchair and zigzag. We will
demonstrate in the subsequent sections that main features in the spin
polarization of the electron density and the enhancement of the $g$-factor
are rather similar for both types of edges. There is however an important
difference between them which can be traced to the presence of the
zero-energy mode (ZEM) residing at the edge of the zigzag GNRs.\cite%
{Wakabayashi} In the present section we will demonstrate that the ZEM leads
to a different features in the potential and charge density profiles near
the edges for the cases of armchair and zigzag ribbons.

Figure \ref{fig:edge} shows the self-consistent charge density distributions
and potential profiles for the armchair and zigzag nanoribbons of various
widths $W$=10, 30, 50 nm. In all calculation the distance between the GNR
and the gate is $d=30$ nm. The calculations are performed in the Hartree
approximation for spinless electrons at zero field (i.e. $V_{U}^{\sigma }$
and $V_{Z}^{\sigma }$ are set to $0$ in the Hamiltonian (\ref{H})). It is
noteworthy that the electron density distribution obtained from the
electrostatics (i.e. due to the Hartree potential $V_{H}(\mathbf{r}),$ Eq. (%
\ref{VH})) is not altered significantly by magnetic field \cite{Chklovskii}.
The charge densities and potentials stay qualitatively the same as the
nanoribbon width increases and exhibit practically no difference for 30- and
50-nm wide nanoribbons. We thus conclude that the transverse confinement
does not change substantially for nanoribbons wider than $\sim 30$ nm, and
the width $W=30$ nm is sufficient to describe realistically wide ribbons or
even an edge of a graphene sheet.

Let us now focus on a difference in the potential profiles and the electron
density distributions in a vicinity of a ribbon edge for armchair and zigzag
ribbons. Both ribbons show strong electron accumulation near the edges but
this accumulation is stronger in the zigzag GNRs. The corresponding
potential profiles for armchair and zigzag ribbons have different shapes
near the edges. For the armchair ribbon, the potential has a triangular
shape, see Fig. \ref{fig:edge}(b). This was predicted and explored
previously.\cite{Silvestrov08,Shylau09} The triangular shape of the
potential is related to the hard-wall confinement. It is noteworthy that a
similar triangular shape of a potential is exhibited by cleaved-edge
overgrown quantum wires where electrons also experience a hard-wall
confinement. \cite{Grayson,IhnatsenkaCEOQW}.

The potential profile for the case of zigzag ribbon exhibits somehow
different features. As in the case of the armchair GNRs, it gradually
decreases towards the boundaries to form a well in the vicinity of the
edges. However, in the close proximity to the edges, it raises up and
crosses the Fermi energy, see Fig. \ref{fig:edge}(d). We relate this feature
to the zero-energy mode (ZEM) that traps charges. The zero-energy mode is
manifested as disperseless energy level pinned to $E_{F}$ in the ranges $%
k\in \frac{\pi }{\sqrt{3}a}\left[ \frac{2}{3},1\right] $ and $k\in -\frac{%
\pi }{\sqrt{3}a}\left[ \frac{2}{3},1\right] $. It is these trapped charges
that raise the potential at the edges. They effectively repulse excess
charges induced in the ribbon by the gate and prevent the triangular well to
form near the boundary. Therefore, the difference in the charge accumulation
and potential profiles in the armchair and zigzag ribbons occurs due to
topological property of the zigzag edge termination supporting the
zero-energy mode. It is important to stress that this difference persists
into the high-field regime and can not be addressed by semi-classical
approaches like in Ref. \onlinecite{Silvestrov08}.

\subsection{LDOS, magnetobandstructure and formation of compressible strips
for spinless electrons}

%*********************************************************
\begin{figure}[tbh]
\includegraphics[keepaspectratio,width=\columnwidth]{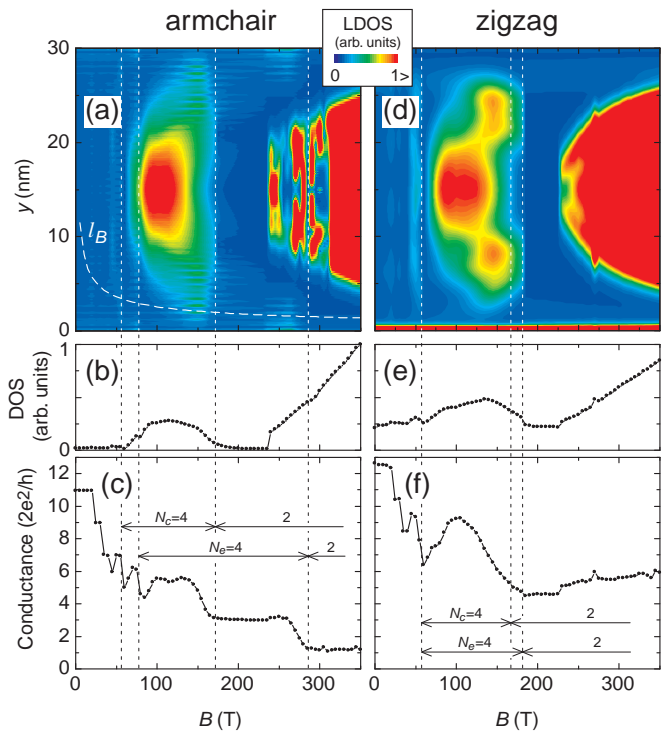}
\caption{(Color online). (a),(d) The LDOS and (b),(e)the DOS at the Fermi
energy, and (c),(f) the two-terminal conductance calculated in the Hartree
approximation for the armchair and zigzag ribbons (left and right panels
respectively). The LDOS is shown for one graphene sublattice, say A; The
LDOS for the sublattice B is symmetric with respect to the ribbon's axis.
The regions with higher LDOS point to higher probability to find an electron
there and correspond to positions of the compressible strip. $N_{c}$ and $%
N_{e}$ in (c) and (f) mark the number of occupied electron subbands in the
center and near the edge of the ribbon respectively. The width of ribbons is
$W=30$ nm corresponding to 242 and 141 carbons in the cross section of
armchair and zigzag ribbons, respectively. $E_{F}$=5 $t$. The temperature $%
T=10$ K.}
\label{fig:DOS}
\end{figure}
%*********************************************************
Spin polarization in conventional quantum wires is related to the formation
of compressible strips\cite{Chklovskii} in the case of interacting spinless
electrons.\cite{Ihnatsenka_wire_comp_strips,IhnatsenkaCEOQW,IhnatsenkaMarcus}
In this section we therefore outline the electronic and transport properties
of armchair and zigzag nanoribbons in the Hartree approximation for spinless
electrons (i.e. disregarding the Hubbard and the Zeeman interactions, $V_{Z}=
$ $V_{U}=0$, in the Hamiltonian Eq. (\ref{H})) focussing on the formation of
the compressible strips.
%*********************************************************
\begin{figure*}[tbh]
\includegraphics[keepaspectratio,width=\textwidth]{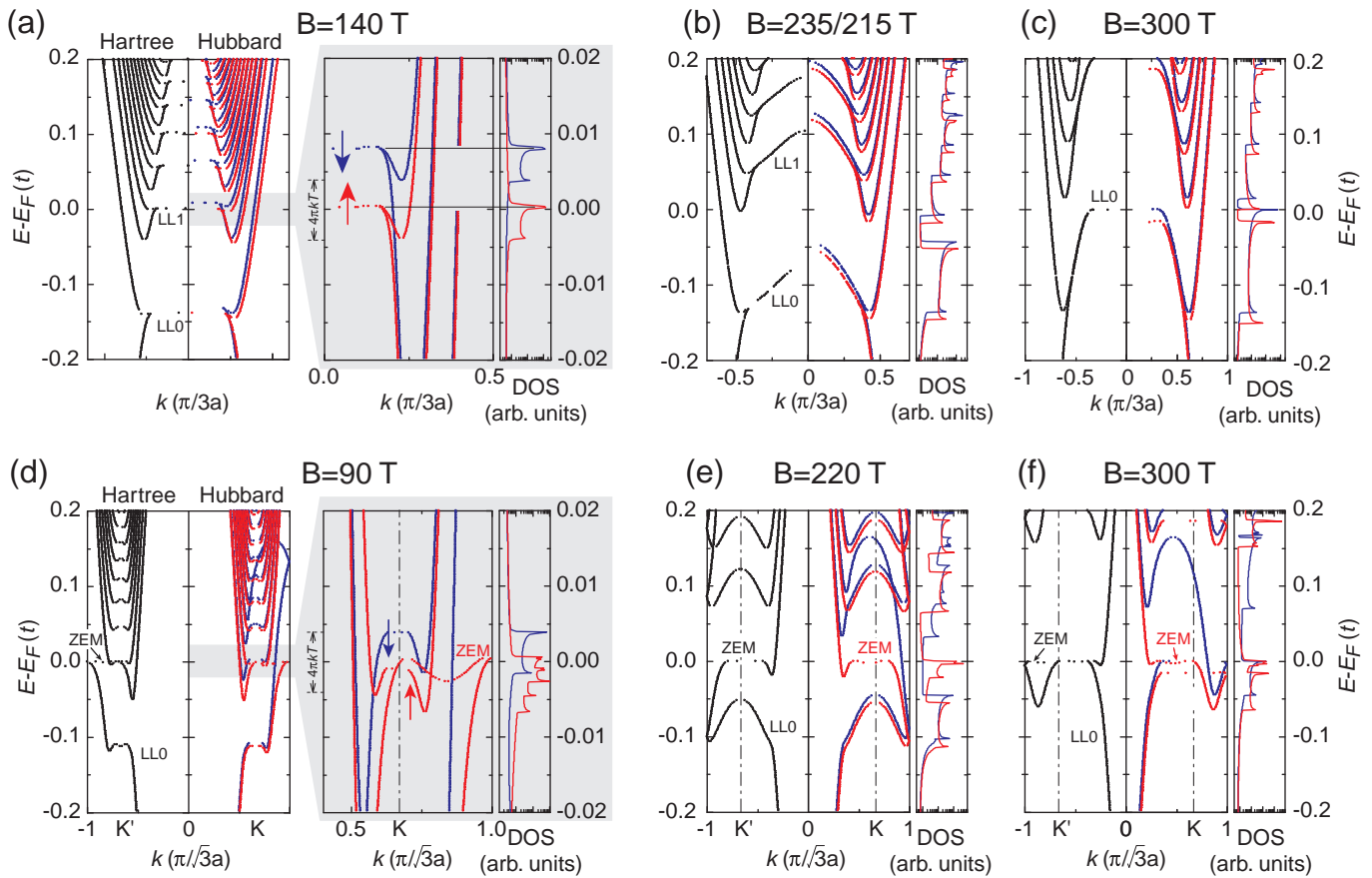}
\caption{(Color online). The dispersion relation for (a)-(c)  the armchair
and (d)-(f) the zigzag ribbons. Left and right panels in every plot
correspond respectively to the Hartree and Hubbard approximations. Because
of the symmetry, only a half of the band diagrams is shown. The red and blue
dotted lines show respectively spin-up and spin-down components. Vertical
dot-dashed lines in (d)-(f) mark $K$ and $K^{\prime }$ points of the first
Brillouin zone of graphene in the zigzag ribbon. In armchair ribbon, $K$ and
$K^{\prime }$ points coincide at $k=0$. In (d)-(f), "ZEM" marks the
zero-energy mode that stays pinned to the Fermi energy.}
\label{fig:band_structure}
\end{figure*}
%*********************************************************

Figures \ref{fig:DOS}(a) (b) show the local density of states (LDOS), $%
LDOS(y)=-\int \rho (y,E)\frac{\partial f_{FD}(E-E_{F})}{\partial E}dE$, and
the total density of states (DOS), $DOS=\sum_{y}LDOS(y)$, at the Fermi
energy as a function of the magnetic field for the armchair and zigzag
ribbons. (Note that LDOS is shown for one sublattice only.) It is noteworthy
that the DOS can be accessible via magneto-capacitance or magnetoresistance
measurements similarly to conventional semiconductor structures defined in
2DEG\cite{Weiss89,Berggren98}. The structure of the LDOS and DOS can be
understood from an analysis of the magnetosubband structure. We outline
below the main features of the subband structure for the armchair and zigzag
ribbons focussing on the differences and similarities between them as well
as on formation of the compressible strips in the ribbons. (Note that
evolution of the band structure for the case of the armchair GNRs was
discussed by Shylau \textit{et al.}\cite{Shylau10}).

Left panels of Figs. \ref{fig:band_structure} (a)-(c) show the band
structure of armchair graphene nanoribbons for spinless Hartree electrons
for three representative magnetic fields. Flat regions in the band diagrams
correspond to the Landau levels in bulk graphene, and dispersiveness states
close to the GNRs boundaries represent edge states corresponding to
classical skipping orbits. Figure \ref{fig:band_structure} (a) shows the
band diagram for a magnetic field $B=140$ T when the two lowest Landau
levels, LL0 and LL1, are filled. The first Landau level LL1 is pinned to the
Fermi energy thus forming a compressible strip in the center of the GNR. The
strip is called compressible when the electron density can be easily
redistributed in order to effectively screen the external potential. We
define a compressible strip as a region where the dispersion lies within the
energy window $|E-E_{F}|\lesssim 2\pi k_{B}T$ \cite%
{Ando,Ihnatsenka_wire1,Ihnatsenka_wire_comp_strips,Shylau10} because in this
energy window the states are partially filled i.e., $0<f_{FD}<1$ and thus
the electron density can be easily changed.

Because a graphene ribbon has abrupt edges the self-consistent potential
forms the triangular wells near edges as discussed in the previous section.
As a result the center of the ribbon and its edges depopulate in magnetic
field differently. Namely, as the magnetic field increases, the subbands
first depopulate in the center and then near the edges. For example, the
second subband (i.e. LL1) is pinned to the Fermi level in the ribbon center
in the interval $B\approx 60-170$ T. Within this interval it forms the
compressible strip which is manifested itself as the enhanced LDOS and DOS
in Figs. \ref{fig:DOS}(a) and (b). However, the LL1 stays populated near the
edges in a wider magnetic field range, $B\approx 80-290$ T . In the field
interval $B\approx 170-230$ T the LL1 is depopulated in the center (see Fig. %
\ref{fig:band_structure} (b)). As a result the LDOS and DOS are practically
zero, see Figs. \ref{fig:DOS} (a),(b). When the magnetic field increases to $%
B\approx 230$ T, the lowest Landau level, LL0, is pushed up in energy and
gets pinned to the Fermi energy. This again leads to a formation of the
compressible strip in the middle of the wire (see Fig. \ref%
{fig:band_structure} (b)) and to the enhancement of the LDOS and DOS at the
Fermi energy as seen in Figs. \ref{fig:DOS} (a),(b). At this $B$, two
different LLs are at $E_{F}$ and contribute to electron transport: LL0 in
the center and LL1 near the edges.

The graphene ribbons with the zigzag edge termination exhibits features of
the magnetosubband structure similar to those for the armchair terminated
ribbons. This is illustrated in Figs. \ref{fig:band_structure} (d)-(f) which
show the band structure of the zigzag GNRs for three representative magnetic
fields, $B=90$ T, 220 T and 300 T. As for the armchair GNRs, these fields
correspond respectively to the cases when the LL1 is pinned to $E_{F}$ in
the middle of the GNR (Fig. \ref{fig:band_structure} (d)), LL1 is
depopulated in the middle of the GNR (Fig. \ref{fig:band_structure} (e)),
and LL0 is pinned to $E_{F}$ in the middle of the GNR (Fig. \ref%
{fig:band_structure} (f)). However, there are several striking differences
between the zigzag and armchair ribbons manifested in their LDOS, DOS and
the subband structure. First, strong electron accumulation and formation of
the compressible strip takes place near the very ribbon's boundaries over
all the range of magnetic fields studied, see enhanced LDOS at $y\sim 0$ in
Fig. \ref{fig:DOS}(d). Because of this the total DOS in the zigzag GNR never
drops to zero over all the range of magnetic fields, see Fig \ref{fig:DOS}%
(e). (Note that Fig. \ref{fig:DOS}(d) shows the LDOS for the sublattice A
which is enhanced at the one edge of the ribbon. The sublattice B has the
enhanced LDOS near the opposite edge of the ribbon).

Inspection of the magnetoband structure reveals that these features are
caused by the ZEM in the zigzag GNR discussed in the previous section. This
mode (marked by "ZEM" in Figs. \ref{fig:band_structure} (d)-(f)) always
stays pinned to the Fermi energy because of the high density of states for
electrons that this mode accommodates. It is important to stress that the
pinning of this mode to $E=E_{F}$ is the result of the electron-electron
interaction, and the pinning effect is apparently absent in the one-electron
description where this mode the is always situated at $E=0$.\cite%
{Wakabayashi}

Figures \ref{fig:DOS}(c),(f) show evolution of the two-terminal conductance
as the magnetic field increases for the armchair and zigzag ribbons. In
contrast to the conventional semiconductor quantum wires and quantum point
contacts exhibiting a step-like conductance\cite{QPC}, the conductance of
GNRs reveals non-monotonic decrease with bumps coexisting with quantized
plateau regions of multiples $\frac{2e^{2}}{h}$. Note that $\mathcal{G}$
changes by two conductance quantum between plateaus due to valley degeneracy
of graphene. The origin of the bumps in the conductance was discussed by
Shylau \textit{et al.} \cite{Shylau10} and was related to the
interaction-induced modifications of the band structure leading to the
formation of compressible strips in the middle of GNRs.

\subsection{Spin splitting and the enhancement of the $g$-factor}

Let us now turn to the case of electrons with spin and analyze how the
Hubbard and the Zeeman interactions modify the magnetosubband structure of
graphene nanoribbons leading to the spin polarization and enhancement of the
$g$-factor. For the Hubbard constant we choose $U=t$ which corresponds to
the estimation of Refs. \cite{Yazyev, Jung}. Note that the recent work\cite%
{WehlingKatsnelson2011} predicts a somehow larger value, $U\approx 3t.$
While the utilization of larger value of $U$ leads to some quantitative
differences with the results presented below, they remain qualitatively the
same and the conclusions are not affected. We also stress that while the
discussion below is focused on the case of high magnetic field when the two
lowest Landau levels, LL1 and LL0, are occupied, the similar conclusion
remain valid for lower fields as well.

%*********************************************************
\begin{figure}[th]
\includegraphics[keepaspectratio,width=\columnwidth]{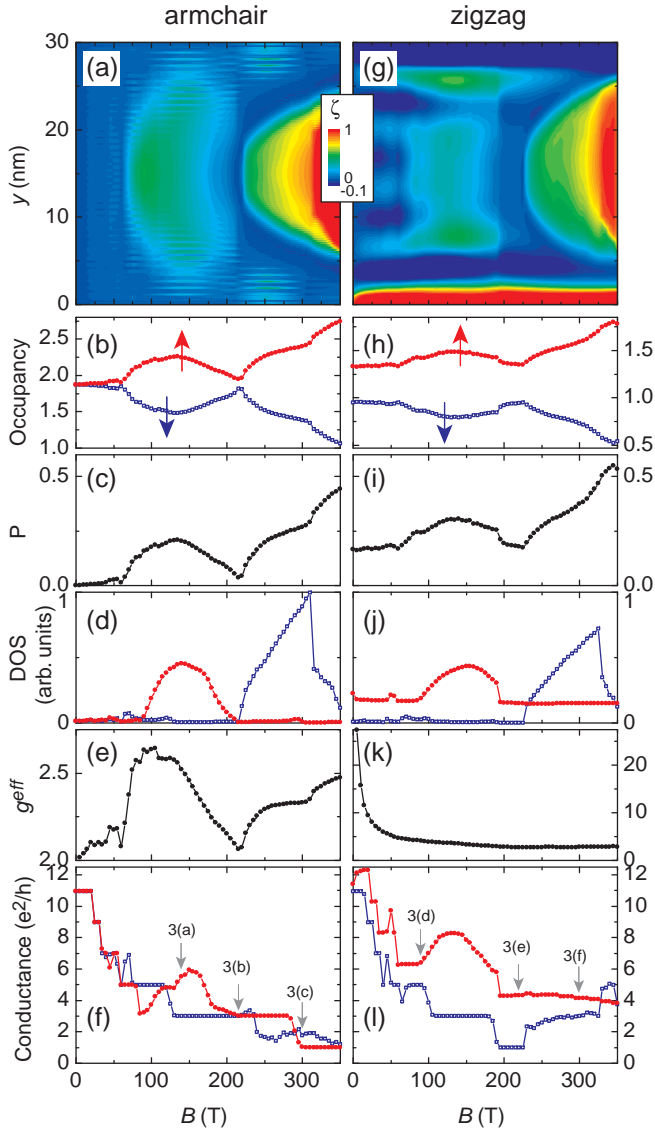}
\caption{(a),(g) The local spin polarization of the charge density, (b),(h)
the unit cell occupancy, (c), (i) the spin polarization $P$, (d),(j) the
DOS, (e), (k) the effective $g$-factor, and (f),(l) the conductance as a
function of magnetic field calculated in the Hubbard approximation for the
armchair (a)-(f) and zigzag (g)-(l) ribbons. The red lines with filled dots
show spin-up component and blue line with open squares is for spin-down.
Arrows in (f) and (l) mark magnetic fields for the band structures shown in
Fig. \protect\ref{fig:band_structure}.}
\label{fig:spin}
\end{figure}
%*********************************************************

Let us start with the case of the armchair ribbons. Figures \ref{fig:spin}
(a)-(c) show the local spin polarization of the charge density, $\zeta (y)=%
\frac{n^{\uparrow }(y)-n^{\downarrow }(y)}{n^{\uparrow }(y)+n^{\downarrow
}(y)},$ spin-resolved densities and the total spin polarization, $P=\frac{%
n^{\uparrow }-n^{\downarrow }}{n^{\uparrow }+n^{\downarrow }},$ ($n^{\sigma
}=\sum_{y}n^{\sigma }(y)$) as a function of magnetic field. The features in $%
\zeta (y)$ show a striking similarity with the features of the LDOS, and the
behavior of the $P$ follows that one of the DOS calculated in the Hartree
approximation (c. f. Fig. \ref{fig:DOS} (a),(b)). This similarity is not
coincidental. The regions with the enhanced LDOS correspond to compressible
strips, and it is the compressible strips where the spin splitting of
subbands takes place. Indeed, in the compressible region the subbands are
only partially filled because $f_{FD}<1$ there, and, therefore, the
population of the spin-up and spin-down subbands can be easily changed. This
population difference triggered by Zeeman splitting is enhanced by the
Hubbard interaction leading to different effective potentials for spin-up
and spin-down electrons and eventually to the subband spin splitting.

For a more detailed analysis let us follow an evolution of the band
structure in Fig. \ref{fig:band_structure}(a)-(c). The right panel of Fig. %
\ref{fig:band_structure} (a) shows a spin-resolved magnetoband structure
corresponding to the case when LL1 forms a compressible strip in the middle
of the ribbon for the case of spinless electrons (c.f. the left panel of the
figure). The Hubbard interaction pushes up the spin-down subbands above the
window $|E-E_{F}|\approx 2k_{B}T$ such that it becomes depopulated, and the
compressible strip in the middle is occupied by spin-up electrons only. As a
result, the DOS at $E_{F}$ of the spin-up electrons is enhanced, while that
one of the spin-down electrons is zero, see Fig. \ref{fig:spin} (d). All
these lead to the difference in the electron densities $n^{\uparrow }$ and $%
n^{\downarrow }$ and the spin polarization in the ribbon as shown in Figs. %
\ref{fig:spin} (b) and (c). When the magnetic field increases such that the
LL1 is pushed above the $E_{F},$ the compressible strip in the middle
disappears, the DOS at $E_{F}$ for both spin species becomes equal to zero
and the spin polarization vanishes. This is illustrated in the band diagram
shown in Fig. \ref{fig:band_structure} (b) corresponding to the case when $%
E_{F}$ is situated between LL0 and LL1. With further increase of the
magnetic field the LL0 is pushed up to $E_{F}.$ As this subband is pushed
from below, in this case it is the higher energy spin-down state that gets
pinned to $E_{F}$ forming a compressible strips, whereas the spin-up subband
in the middle of the ribbon remains below the $|E-E_{F}|\approx 2k_{B}T.$ As
a result, for this case the DOS at $E_{F}$ for the spin-down electrons is
larger than that for the spin-up electrons, Fig. \ref{fig:spin}(d). Note
that despite of this, $n^{\uparrow }>n^{\downarrow }$ (see Fig. \ref%
{fig:spin} (b)) because the spin-up subband is fully occupied, whereas the
spin-down subband is occupied only partially.

Figure \ref{fig:spin}(e) shows the effective $g^{\ast }$-factor for the
armchair ribbon defined according to $g^{\ast }=\langle V^{\uparrow
}-V^{\downarrow }\rangle /\mu _{B}B$, where the averaging is done over all
carbon atoms. %(is it the same as ($\langle
%V^{\uparrow }\rangle -\langle V^{\downarrow }\rangle )/\mu _{B}B? )$ {\color{red}{Yes}}.
Because the electron density is related to the potential, the features in
the $g^{\ast }$-factor resemble those of the polarization $P,$ showing the
behavior reflecting successive population and depopulation of the spin-up
and spin-down subbands. %The effective $g^{\ast }$-factor varies around a
%value of $\left\langle g^{\ast }\right\rangle \approx 2.25,$which represents
%a rather modest enhancement $\sim 12\%$ in comparison to the case of
%noninteracting electrons in pristine graphene with $g=2.$Note that in bulk
%graphene the effective $g^{\ast }$-factor is reported to be $g^{\ast
%}\approx 2.7$. (Ref.)

In high magnetic field, $B\gtrsim 50$ T, (corresponding to population of LL1
and LL0) the effective $g$-factor varies between $2.1\lesssim g^{\ast }\leq
2.7$ with the average value $\left\langle g^{\ast }\right\rangle \approx 2.25
$, which represents a rather modest enhancement in comparison to the case of
noninteracting electrons in pristine graphene with $g=2$. Note that in the
bulk graphene the effective $g^{\ast }$-factor was reported to be $g^{\ast
}\approx 2.7.$ \cite{Kurganova}

It is worth mentioning that the main features of the spin polarization and
the subband evolution in magnetic field resemble those in the cleaved-edge
overgrown quantum wires (CEOQW) \cite{IhnatsenkaCEOQW}. This is because in
both cases the potential corresponds to the hard-wall confinement. The
difference is that in CEOQW, as well as in conventional GaAs split-gate wires%
\cite{Ihnatsenka_wire1}, the polarization and thus the effective $g^{\ast }$%
-factor are enhanced by a factor of $\sim 10$ in comparison to the Zeeman
splitting, whereas in the armchair ribbons this enhancement it is just $\sim$
0.22. This can be explained by fact that the bare $g$-factor in armchair
GNRs is much larger that that one in GaAs ( $g_{\text{GaAs}}/g_{\text{%
graphene}}\approx 0.022$), such that Zeeman interaction in graphene remains
dominant in comparison to the exchange one.

One more important difference of the graphene ribbons from the conventional
GaAs quantum wires is in the character of spin polarized edge states in the
vicinity of the boundaries. In the conventional quantum wires the edge state
of opposite spins are spatially separated.\cite{Ihnatsenka_wire_comp_strips}
This is ultimately related to the formation of the compressible strips \emph{%
near the boundaries} of the split-gate wire because of the soft confinement
due to the gates.\cite{Ihnatsenka_wire_comp_strips} In GNR due to the
hard-wall confinement the compressible strips do not form near the
boundaries, and hence the spatial separation of the edge states of opposite
spins does not occur. It is worth mentioning that in this respect the GNRs
are also similar to CEOQW.\cite{IhnatsenkaCEOQW}

Let us now turn to the case of zigzag nanoribbons. The main features of the
spin polarization and the subband evolution are rather similar to the
armchair ones. There is however one important difference related to the
presence of the zero-energy mode residing at the zigzag edges. In contrast
to other modes exhibiting successive population and depopulation of the
Landau levels in the middle of the ribbon, this mode always stays pinned to $%
E_{F}$ thus forming a compressible strip with the enhanced density of states
at the edges for all magnetic fields . The Hubbard interaction leads to a
complete spin polarization such that electrons in this mode are always in
the spin-up state. This is seen in the spatially-resolved polarization shown
in Fig. \ref{fig:spin}(g).
%(Note that Fig. \ref{fig:spin}(?) shows the polarization for the sublattice A only. The polarization for the sublattice B (not shown here) is symmetric with respect to the ribbon's axis).
Because of this the total spin-up density is significantly larger than the
spin-down one for all magnetic fields, and the DOS for spin-up electrons
never drops to zero. This is in contrast to the case of armchair GNRs where
electron densities for opposite spin species can be equal when the Fermi
energy lies between two consecutive Landau levels, and where both DOS for
spin-ups and spin-downs can drop to zero (c.f. Figs. (d) with (j)). Because
of the strong polarization near zigzag edges, the effective $g^{\ast }$%
-factor is strongly enhanced at low field (when the Zeeman spliting is
small), and at higher fields it decreases to values $g^{\ast }\approx 3$
comparable to those in the armchair GNRs, see Fig. \ref{fig:spin} (k).

Figures \ref{fig:spin}(f),(l) show evolution of the two-terminal conductance
as the magnetic field increases for the armchair and zigzag ribbons. The
conductance is apparently spin polarized with $\mathcal{G}^{\uparrow }\neq $
$\mathcal{G}^{\downarrow }$. This reflects the fact that at a given magnetic
field the number of propagating states at $E_{F}$ accommodating spin-up and
spin-down electrons are different. As in the case of spinless electrons, the
spin-resolved conductance also exhibits a bump-like structure which origin
is the same as for the case of spinless electrons. Note that because of the
presence of a spin-polarized zero-energy mode, the spin-up conductance is
larger than spin-down one for most fields. Finally we stress that Figs. \ref%
{fig:spin}(f),(l) show the conductances of ideal GNRs without defects. The
defects scattering will modify GNRs conductance, especially for the
low-velocity modes flowing in the middle of the ribbons. At the same time,
the edge states corresponding to the classical skipping orbits are robust
against the impurity scattering.\cite{Shylau10}

Finally we note that we also performed similar computations of the spin
polarization for GNRs using the density functional theory with the exchange
functional proposed by Polini \textit{et al}.\cite{Polini} including the
spin degree of freedom as prescribed in Ref. \onlinecite{GiulianiVignale}.
Practically no spin polarization was observed that we attribute to the
positive sign of the exchange energy in Ref. \onlinecite{Polini}. More
systematic studies of the spin polarization in GNRs using different approach
(such as the spin-DFT, Hartree-Fock, etc.) would be very interesting.

Note that spin polarization and the enhancement of the g-factor in bulk
graphene in the presence of impurities was recently studied by Volkov
\textit{et al.} \cite{Anton}.

\section{Conclusions}

We provide a systematic quantitative description of the spin polarization,
the subband structure and the density and potential profiles in the armchair
and zigzag graphene nanoribbons in a perpendicular magnetic field. In our
study we addressed realistically wide nanoribbons, and our conclusion
concerning the density and potential distributions near the edge can also be
applied for the case of a semi-infinite graphene sheet. Our calculations are
based on the self-consistent Green's function technique where electron
interaction and spin effects are included by the Hartree and the Hubbard
potentials.

We first focus on the case of spinless electrons and find that the potential
profile and the density distribution are different near the edges of the
armchair and zigzag ribbons. For the armchair termination, the potential at
the edge has a triangular shape whereas for the zigzag ribbons it exhibits a
well-type character. Both terminations show strong electron accumulation
near the edges but this accumulation is stronger at the the zigzag edge.
This difference is attributed to a topological property of the zigzag edge
termination supporting the zero-energy mode.

Because of the spin polarization in nanoribbons and conventional quantum
wires is ultimately related to the formation of the compressible strips for
the case of spinless electrons, we study the LDOS, DOS, and the
magnetosubband structure for the armchair and zigzag ribbons focussing on
the differences and similarities between them as well as on the formation of
the compressible strips in the ribbons. For both types of nanoribbons we
find that a compressible strip with the enhanced DOS forms in the middle of
the ribbon in accordance with the successive population and depopulation of
the Landau levels. For the case of zigzag edge termination we find a strong
electron accumulation and formation of a compressible strip near the edges
over all the range of magnetic fields. This is caused by the presence of the
zero-energy mode that always stays pinned to the Fermi energy because of the
high DOS that this mode accommodates.

Accounting for the Zeeman interaction and describing the spin effects via
the Hubbard potential we discuss how the spin-resolved subband structure
evolves when an applied magnetic field varies. We find that the local spin
polarization of the electron density and the total spin polarization exhibit
a behavior similar to that one of the LDOS and DOS for spinless electrons.
This similarity is not coincidental and reflects the fact that the regions
with the enhanced DOS correspond to compressible strips where the spin
splitting of subbands takes place. We find that for the armchair ribbons in
high magnetic field the effective $g$-factor varies between $2.1\lesssim
g^{\ast }\leq 2.7$ with the average value $\left\langle g^{\ast
}\right\rangle \approx 2.25$. For the zigzag nanoribbons we find that the
zero-energy mode remains pinned to $E_{F}$ and becomes fully spin-polarized
for all magnetic fields, which, in turn, leads to a strong spin polarization
of the electron density near the zigzag edge. Due to the contribution of the
fully spin polarized zero-energy mode the effective $g^{\ast }$-factor in
the armchair GNRs is strongly enhanced at low field (when the Zeeman
splitting is small), and at higher fields it decreases to values $g^{\ast
}\approx 3$ comparable to those in the armchair GNRs.

It is worth mentioning that the main features of the spin polarization and
the subband evolution in magnetic field resemble those in the cleaved-edge
overgrown quantum wires (CEOQW). This is because in both cases the potential
corresponds to the hard-wall confinement.

Finally, we stress the importance of accounting for the global
electrostatics in the system at hand for the accurate description of the
spin polarization in GNRs. (In the present study it is done by accounting
for the long-range Coulomb interaction by means of the self-consistent
Hartree potential). This is because the global electrostatic is responsible
for the formation of the compressible strips, and it is the compressible
strips where the spin splitting of subbands takes place.

\begin{acknowledgments}
Authors acknowledge a collaborative grant from the Swedish Institute.
\end{acknowledgments}


\begin{thebibliography}{99}
\bibitem{Castro_Neto_review} A. H. Castro Neto, F. Guinea, N. M. R. Peres,
K. S. Novoselov and A. K. Geim, Rev. Mod. Phys. \textbf{81}, 109 (2009).

\bibitem{GusyninQHE} V. P. Gusynin, and S. G. Sharapov, Phys. Rev. Lett.
\textbf{95}, 146801 (2005).

\bibitem{Novoselov2005} K. S. Novoselov, A. K. Geim, S. V. Morozov, D.
Jiang, M. I. Katsnelson, I. V. Grigorieva, S. V. Dubonos, and A. A. Firsov,
Nature (London) \textbf{438}, 197 (2005).

\bibitem{Zhang} Y. Zhang, Y.-W. Tan, H. L. Stormer, and P. Kim, Nature
(London) \textbf{438}, 201 (2005).

\bibitem{Zhang2006} Y. Zhang, Z. Jiang, J. P. Small, M. S. Purewal, Y.-W.
Tan, M. Fazlollahi, J. D. Chudow, J. A. Jaszczak, H. L. Stormer, and P. Kim,
Phys. Rev. Lett. \textbf{96}, 136806 (2006).

\bibitem{Jiang} Z. Jiang, Y. Zhang, H. L. Stormer, and P. Kim, Phys. Rev.
Lett. \textbf{99}, 106802 (2007).

\bibitem{Checkelsky} J. G. Checkelsky, L. Li, and N. P. Ong, Phys. Rev.
Lett. \textbf{100}, 206801 (2008).

\bibitem{Zhang2009} L. Zhang, J. Camacho, H. Cao, Y. P. Chen, M. Khodas, D.
E. Kharzeev, A. M. Tsvelik, T. Valla, and I. A. Zaliznyak, Phys. Rev. B
\textbf{80}, 241412(R) (2009).

\bibitem{Zhao12} Yue Zhao, Paul Cadden-Zimansky, Fereshte Ghahari, and
Philip Kim, Phys. Rev. Lett. \textbf{108}, 106804 (2012).

\bibitem{Kurganova} E. V. Kurganova, H. J. van Elferen, A. McCollam, L. A.
Ponomarenko, K. S. Novoselov, A. Veligura, B. J. van Wees, J. C. Maan, and
U. Zeitler, Phys. Rev. B \textbf{84}, 121407(R) (2011).

\bibitem{Folk} M. B. Lundeberg and J. A. Folk, Nature Phys. 5, 894 (2009).

\bibitem{Guttinger} J. G\"{u}ttinger, T. Frey, C. Stampfer, T. Ihn, and K.
Ensslin, Phys. Rev. Lett. \textbf{105}, 116801 (2010).

\bibitem{Kinaret} J. M. Kinaret and P. A. Lee, Phys. Rev. B \textbf{42},
11768 (1990).

\bibitem{Dempsey} J. Dempsey, B. Y. Gelfand, and B. I. Halperin, Phys. Rev.
Lett. \textbf{70}, 3639 (1993).

\bibitem{Tokura} Y. Tokura and S. Tarucha, Phys. Rev. B \textbf{50}, 10981
(1994).

\bibitem{takis2002} Z. Zhang and P. Vasilopoulos, Phys. Rev. B \textbf{66},
205322 (2002).

\bibitem{Stoof} T. H. Stoof and G. E. W. Bauer, Phys. Rev. B \textbf{52},
12143 (1995).

\bibitem{Ihnatsenka_wire1} S. Ihnatsenka and I. V. Zozoulenko, Phys. Rev. B
\textbf{73}, 075331 (2006).

\bibitem{Ihnatsenka_wire_comp_strips} S. Ihnatsenka and I. V. Zozoulenko,
Phys. Rev. B \textbf{73}, 155314 (2006).

\bibitem{IhnatsenkaCEOQW} S. Ihnatsenka and I. V. Zozoulenko, Phys. Rev. B
\textbf{74}, 075320 (2006).

\bibitem{IhnatsenkaMarcus} S. Ihnatsenka and I. V. Zozoulenko, Phys. Rev. B
\textbf{78}, 035340 (2008).

\bibitem{spintronics} I. \v{Z}uti\'{c}, J. Fabian, and S. Das Sarma, Rev.
Mod. Phys. 76, 323 (2004).

\bibitem{adot} I. V. Zozoulenko and M. Evaldsson, Appl. Phys. Lett. \textbf{%
85}, 3136 (2004).

\bibitem{Giovannetti} B. Karmakar, D. Venturelli, L. Chirolli, F. Taddei, V.
Giovannetti, R. Fazio, S. Roddaro, G. Biasiol, L. Sorba, V. Pellegrini, and
F. Beltram, Phys. Rev. Lett. \textbf{107}, 236804 (2011).

\bibitem{Camino} F. E. Camino, Wei Zhou, and V. J. Goldman, Phys. Rev. B
\textbf{72}, 155313 (2005).

\bibitem{IhnatsenkaKirzcenow} S. Ihnatsenka, I. V. Zozoulenko, and G.
Kirczenow, Phys. Rev. B \textbf{80}, 115303 (2009).

\bibitem{Paradiso} N. Paradiso, S. Heun, S. Roddaro, D. Venturelli, F.
Taddei, V. Giovannetti, R. Fazio, G. Biasiol, L. Sorba, and F. Beltram,
Phys. Rev. B \textbf{83}, 155305 (2011).

\bibitem{Palacios} J. Fern\'{a}ndez-Rossier and J. J. Palacios, Phys. Rev.
Lett. \textbf{99}, 177204 (2007).

\bibitem{Yazyev} O. V. Yazyev, Phys. Rev. Lett. \textbf{101}, 037203 (2008).

\bibitem{Xu} H. Xu, T. Heinzel, M. Evaldsson, and I. V. Zozoulenko, Phys.
Rev. B \textbf{77}, 245401 (2008).

\bibitem{Shylau10} A. A. Shylau, I. V. Zozoulenko, H. Xu, and T. Heinzel,
Phys. Rev. B \textbf{82}, 121410(R) (2010).

\bibitem{Shylau11} A. A. Shylau and I. V. Zozoulenko, Phys. Rev. B \textbf{84%
}, 075407 (2011).

\bibitem{Molitor} F. Molitor, A. Jacobsen, C. Stampfer, J. G\"{u}ttinger, T.
Ihn, and K. Ensslin, Phys. Rev. B \textbf{79}, 075426 (2009).

\bibitem{Oostinga} J. B. Oostinga, B. Sacepe, M. F. Craciun, and A. F.
Morpurgo, Phys. Rev. B \textbf{81}, 193408 (2010).

\bibitem{Poumirol} J.-M. Poumirol, A. Cresti, S. Roche, W. Escoffier, M.
Goiran, X. Wang, X. Li, H. Dai, and B. Raquet, Phys. Rev. B 82, 041413(R)
(2010).

\bibitem{Ribeiro} R. Ribeiro, J.-M. Poumirol, A. Cresti, W. Escoffier, M.
Goiran, J.-M. Broto, S. Roche, and B. Raquet, Phys. Rev. Lett., \textbf{107}
086601 (2011)

\bibitem{Wakabayashi} K. Wakabayashi, M. Fujita, H. Ajiki, and M. Sigrist,
Phys. Rev. B \textbf{59}, 8271 (1999).

\bibitem{Chklovskii} D. B. Chklovskii, B. I. Shklovskii, and L. I. Glazman,
Phys. Rev. B \textbf{46}, 4026 (1992); D. B. Chklovskii, K. A. Matveev, and
B. I. Shklovskii, \textit{ibid.} \textbf{47}, 12605 (1993).

\bibitem{Silvestrov08} P. G. Silvestrov and K. B. Efetov, Phys. Rev. B
\textbf{77}, 155436 (2008).

\bibitem{Shylau09} A. A. Shylau, J. W. Klos, and I.V. Zozoulenko, Phys. Rev.
B \textbf{80}, 205402 (2009).

\bibitem{Grayson} M. Huber, M. Grayson, M. Rother, W. Biberacher, W.
Wegscheider, and G. Abstreiter, Phys. Rev. Lett. \textbf{94}, 016805 (2005).

\bibitem{Weiss89} D. Weiss, C. Zhang, R. R. Gerhardts, K. v. Klitzing, G.
Weimann, Phys. Rev. B \textbf{39}, 13020 (1989).

\bibitem{Berggren98} K.-F. Berggren, G. Roos, and H. van Houten, Phys. Rev.
B \textbf{37}, 10118 (1988).

\bibitem{Ando} T. Suzuki and T. Ando, Physica B \textbf{249-251}, 415 (1998).

\bibitem{QPC} C.W. J. Beenakker and H. van Houten, Solid State Physics
Academic, New York, 1991 , Vol. 44, p. 1.

\bibitem{Jung} J. Jung and A. H. MacDonald, Phys. Rev. B \textbf{80}, 235417 (2009).

\bibitem{WehlingKatsnelson2011} T. O. Wehling, E. \c{S}a\c{s}\i o\u{g}lu, C.
Friedrich, A. I. Lichtenstein, M. I. Katsnelson, and S. Bl\"{u}gel, Phys.
Rev. Lett., 106, 236805 (2011)

\bibitem{Polini} M. Polini, A. Tomadin, R. Asgari, and A. H. MacDonald,
Phys. Rev. B \textbf{78}, 115426 (2008).

\bibitem{GiulianiVignale} G. F. Giuliani and G. Vignale, \textit{Quantum
Theory of the Electron Liquid} (Cambridge University Press, Cambridge, 2005).

\bibitem{Anton} A. Volkov, A. A. Shylau, and I. V. Zozoulenko, to be
published.
\end{thebibliography}
\end{document}